\documentclass[utf8]{FrontiersinHarvard} 

\usepackage{url,hyperref,microtype,subcaption}
\usepackage[onehalfspacing]{setspace}
\usepackage{array, multirow}

\makeatletter
\newcommand{\owntag}[2][\relax]{
  \ifx#1\relax\relax\def\owntag@name{#2}\else\def\owntag@name{#1}\fi
  \refstepcounter{equation}\tag{Model \theequation, #2}%
  \expandafter\ltx@label\expandafter{eq:\owntag@name}%
  \edef\@currentlabel{\theequation, #2}\expandafter\ltx@label\expandafter{Eq:\owntag@name}%
  \def\@currentlabel{#2}\expandafter\ltx@label\expandafter{tag:\owntag@name}%
}
\makeatother

\def\keyFont{\fontsize{8}{11}\helveticabold }
\def\firstAuthorLast{Lopez and Bliss}
\def\Authors{Michael J. Lopez\,$^{1,*}$ and Thompson Bliss\,$^{2}$}

\def\Address{$^{1}$National Football League, New York, USA \\
$^{2}$National Football League, New York, USA }

\def\corrAuthor{Michael Lopez}

\def\corrEmail{Michael.Lopez@nfl.com}

\begin{document}
\onecolumn
\firstpage{1}

\title[NFL rest impact]{Bye-Bye, Bye Advantage: Estimating the competitive impact of rest differential in the National Football League} 

\author[\firstAuthorLast ]{\Authors} 
\address{} 
\correspondance{} 

\extraAuth{}
\maketitle

\begin{abstract}

\section{}

The National Football League (NFL) sets its regular season schedule to optimize viewership and minimize competitive inequities. One inequity assumed to impact team performance is rest differential, defined as the relative number of days between games. Using Bayesian state space models on both game outcomes and betting market data, we estimate the competitive effect of rest differential in American football. We find that the most commonly referred to inequities -- both the bye week rest advantage and the mini-bye week rest advantage -- currently show no significant evidence of providing the rested team a competitive edge. Further, we trace a decline in the advantage of a bye week to a 2011 change to the NFL's Collective Bargaining Agreement, which represents a natural experiment to test the relevance of rest and preparation in football. Prior to the agreement, NFL teams off a bye week received a significant advantage (+2.2 points per game), but since 2011, that benefit has been mitigated.

\tiny
 \keyFont{ \section{Keywords:} Bayesian modeling, NFL, rest differential, state space, natural experiment} 
\end{abstract}

\section{Introduction}

\indent In 2023, 93 of the top 100 United States television broadcasts were National Football League (NFL) games, with 83 stemming from the league's regular season \citep{sched93}. The schedule for this regular season is set in the spring, and is the culmination of a months-long process that analyzes, among other factors, rest and travel, stadium availability, and team interest. 

One challenge in setting the schedule each year is the myriad possibilities. Even though an individual team currently plays just 17 games each regular season, those games can be assigned in roughly one billion ways. For all 32 teams, roughly a quadrillion combinations are possible \citep{scheduleqa}. 

The NFL reduces the population of potential schedules by eliminating games and strings of games that are perceived as too unfair \citep{karwan2015alleviating, sched0}. One factor examined in this process is rest differential, broadly defined as the comparison of the relative number of days in between games among two competing teams. Differences in rest could impact, among other factors, recovery time, preparation, injury risk, and travel. In a league with only 17 games, where one win can be the difference between making the playoffs and staying at home, giving one team an unbalanced set of rest differentials could be considered an unfair advantage. As a result, media and fans frequently track how their favorite teams' rest differential stacks up each year \citep{sched1, sched2, sched3}.  

NFL games are most commonly played on Thursday (usually one game, Thursday Night Football, or TNF), Sunday (several games, usually split between 1:00 EST, 4:00 EST, and night), and Monday (usually one game, Monday Night Football, or MNF).\footnote{Additional gamedays can include major American holidays like Thanksgiving and Christmas, as well as unique one-off games or sets of games on Friday's and Saturday's.} If both teams played the week before, a given team could enter with a rest differential of plus or minus four days (one team played Monday, the other Thursday), plus or minus three days (Thursday, Sunday), plus or minus one day (Sunday, Monday), or no rest difference (teams played the same day). But in addition to the weekly differences, each team receives one bye week each regular season. This results in rest differentials as large as plus or minus 8 days. 

The largest perceived advantage is the bye-week benefit (+6 to +8 days). However, league guidelines have varied over the last two decades whether or not teams can practice during that time period. Specifically, the 2011 Collective Bargaining Agreement (CBA) eliminated practice time during bye weeks by guaranteeing four days off for players \citep{nflcba}. What had been two potential weeks of practice for a game before 2011 turned into one week of practice after 2011. As a result, 2011 represents an inflection point for us to test the competitive impact of the bye. In other words, does body rest and time off provide players a rest differential benefit in the NFL? Or is it the extra practice time that, prior to 2011, came alongside bye weeks?

Arguably, the perception of a rest differential impact in the NFL outweighs the evidence. Both \cite{sung2014national} and \cite{sung2020national} found that the bye week provided a competitive advantage that had not yet been picked up in betting markets. However, in neither paper was there a comparison pre and post CBA. Post 2011, \cite{murray2018examining} identified that home teams with a one-day MNF disadvantage lost more often than expected, while finding no benefit to a bye advantage. Finally, \cite{karwan2021analyzing} tied the 2011 CBA change to a decrease in the impact of rest for the home team. However, relative rest was not explicitly modeled. Additionally, \cite{karwan2021analyzing} used the betting market point spread as an explanatory variable in models of point differential, which would conceal the effect of rest differential if betting markets already accounted for it. 

The most well known advocacy's for the competitive impact of rest differences are \cite{sharp23} and \cite{sharp24}. However, these analyses are ripe with small sample sizes, anecdotal evidence (e.g, identifying an under-performing team with negative rest differential), selective endpoints (e.g, only including certain weeks of the season or team types), and the over-interpretation of non-significant results. 

The negative competitive impacts of rest differential are more well established in the National Basketball Association (NBA) (as examples, see \cite{esteves2021basketball, yang2021influence, charest2021impacts, cook2022associations, bowman2023schedule}).  In fact, the inadequate rest associated with the NBA schedule had several experts wanting the league to better balance player sleep as it sets its schedule \citep{singh2021urgent}. In an interview in June of 2024, NBA executive Evan Wasch was direct, stating `we do see a performance decline in the second night of back-to-back games' \citep{nbarest}. Similarly, National Hockey League teams with worst rest differential scored fewer standings points \citep{hockey1, hockey2}, with part of the drop arguably due to a drop in save percentages for goalies playing on zero days rest \citep{hockey3}. 

In professional soccer, \cite{scoppa2015fatigue} found that in the current structure of national team tournaments, different days of rest were not correlated with performance. Related, \cite{watanabe2017weather} found that rest differential was not a significant predictor of team performance in the 2014 World Cup. 

Altogether, there is limited research that directly ties NFL game outcomes to rest differential, and what does exist tends to ignore the 2011 CBA change. Additionally, there are no known looks at betting market odds with respect to the relative point value of rest differential. Betting market numbers offer an important alternative to using game outcomes; in expectation, the point spread will account for all publicly known information prior to a contest, including team strength, game location, and relative rest. As a result, across sport and time, the efficiency of betting market odds is well established (as examples, see \cite{harville1980}, \cite{lacey1990estimation}, \cite{lopez2015building}). Thus, we posit that betting markets will yield more precise estimates of team strength in order to isolate rest differential impact. Finally, simple inefficiencies in betting markets are unlikely to last \citep{vandenbruaene2022efficient}, making it more likely than not that modern NFL betting markets incorporate rest differential into their numbers.

The aim of our paper is to estimate the impact of rest differential in the NFL using modern statistical approaches that can account for team strength. Additionally, we assess if the bye rest benefit changed after the 2011 CBA. We use Bayesian state-space models with time-varying team strength estimates \citep{glickman2017, lopez2018} on two outcomes: point differential and point spread. Our findings suggest that the bye-week advantage existed prior to 2011 (+2.2 points per game), but that most of the bye-week benefit was mitigated after the 2011 CBA (best estimate since 2011, +0.3 points per game). Additionally, while the mini-bye advantage offers no obvious benefit in either game or betting market outcomes, betting markets believe that there is a small advantage to facing an opponent who played the previous Monday night. 

Our paper is laid out as follows. Section \ref{sec:2} proposes state space models of rest differential, while Section \ref{sec:3} details our results. Section \ref{sec:4} concludes and extrapolates the findings to other sports. 

\section{Materials and Methods} \label{sec:2}

\subsection{Methods} \label{sec:3}

The NFL disproportionately allocates prime time slots each season based on which teams are most likely to compete for a championship. Stronger teams receive as many as five or six prime time slots, while lesser teams may only get one or two.\footnote{For example, the 2023 Super Bowl participants have a combined 11 prime time games in the 2024 regular season (Kansas City 5, San Francisco 6), while the Carolina Panthers (who finished with a record of 2-15 in 2023) have 0.} As a result, assessing the impact of rest on NFL game outcomes requires accounting for team strength. Additionally, not all rest discrepancies are treated equivalently for the home and away teams; for example, roughly 60 percent of teams playing the week after a Monday Night Football game play at home. Thus, any estimate of rest differential also requires accounting for the home advantage.

State-space models (\cite{glickman1998, glickman2017, lopez2018}) incorporate time-varying estimates of team strength in statistical models of game outcomes, while likewise accounting for the home advantage. These models are extensions of the more well-known paired comparison models \citep{bradley1952rank}. Using simulations, \cite{benz2023estimating} identified paired comparison models as the framework most robust for estimating home advantages, suggesting that those models would likewise be appropriate for modeling rest advantages. 

In the sections below, we use the state space framework to model game outcomes as a function of team strength, the home advantage, and various categories of rest differential. 

\subsubsection{State Space Models of team strength}

Let $Y_{s,ij}$ be the point differential of a contest between home team $i$ and away team $j$ during season $s$. In our notation, $i, j = 1, ... 32$, and represent each of the NFL's current 32 franchises. We use season $s$, for $s = 2002, ... 2023$. 

The mean outcome $E[Y_{s,ij}]$ between $i$ and $j$ during season $s$ is defined broadly as
\begin{align*}
E[Y_{s,ij}] = \theta_{s,i} - \theta_{s,j} + \mu_{HA, s},
\end{align*}

\noindent where $\theta_{s,i}$ and $\theta_{s,j}$ represent team strength parameters for teams $i$ and $j$ in season $s$, and $\mu_{HA,s}$ is our home advantage specification, such that $\mu_{HA,s} = (\alpha_{HA\ Trend} \times s + \alpha_{HA\ Intercept}) \times I(HA)$. Here, home advantage is treated linearly, given the results of \cite{benz2024comprehensive}, who found a linear home advantage trend provided a better fit relative to constant and categorical alternatives. In this notation, $I(HA)$ is an indicator for if home team $i$ actually has a home advantage. From 2002-2023, 65 games were played at neutral sites, due to either international games or bad weather. 

The team strength parameters, $\theta_{s,i}$ and $\theta_{s,j}$, reflect absolute measures of team ability, and translate into points above a league average team for $i$ and $j$, respectively. Both measures vary stochastically by season. As in \cite{glickman1998} and \cite{lopez2018}, we assume that, in expectation, team strength parameters are pulled towards 0 over time, where, for $s \geq 2003$ and $i = 1, ... 32$,
\begin{align*}
\theta_{(s,i)} \sim N(\gamma \times \theta_{(s-1,i)}, \sigma_{\text{teamstrength}}^2). 
\end{align*}
In this specification, $\gamma$ is an autoregressive parameter for season-to-season variation, and $\sigma_{\text{teamstrength}}^2$ is the variability in team strength. 


\subsubsection{State Space Models of rest difference}

With 4 to 15 days off between each game for each of the home and visiting teams, there are a multitude of potential rest specifications. We propose three categories -- $MNF$, the Monday Night Football advantage, $Mini$, the mini-bye advantage, and $Bye$, the advantage most often linked to a team playing off a bye week - in order to build models to estimate rest effects. Our full split of games into these categories is shown in Table \ref{tab:daysrest}. 

\begin{center}
    (Table \ref{tab:daysrest} here)
\end{center} 

First, our $MNF$ advantage refers to games in which (i) there's at least a one day rest differential between the two teams and (ii) the disadvantaged team is playing on six days rest or less. Usually, the $MNF$ advantage refers to a Sunday game, where one team played the prior Sunday and the other played the prior Monday. Additionally, there other are games where this $MNF$ definition is met, including Saturday to Saturday vs. Sunday to Saturday matchups. 

Second, our $Mini$ advantage refers to games in which both (i) exactly one team had either 9, 10, or 11 days rest and (ii) there was at least a two day difference in rest between the two teams. Usually, the $Mini$ advantage refers to teams playing on Thursday Night Football who return 10 days later to face an opponent playing on seven days rest. There are additional scenarios noted in Table \ref{tab:daysrest}. 

Finally, the $Bye$ advantage refers to teams that did not play the previous week against an opponent that did play the previous week. Generally, these differences are +6 to +8 days. In all situations, these teams are coming off the one regular season bye week given to each NFL team. 

There are a handful of games under unique rest scenarios that we drop from our analysis. These are largely rescheduled games (e.g, COVID in 2020, or games postponed because of weather), many of which were played mid-week, thus creating rare rest discrepancies. More importantly, these settings reflect other disruptions for teams, including illness and/or travel, thus making it difficult to extrapolate to the typical NFL calendar. 

As a final point, we prefer categorical specifications of rest to continuous ones for several reasons. First, the potential impact of rest difference is directly tied to practice and recovery time. As an example, the difference between 13 and 14 days rest (+1 day) is less likely to be impactful than the difference between six and seven days rest (also +1 day). Within each of our three specifications, there is similarity in practice time, days off, and travel, an assertion not always true among teams with the same difference in rest days alone. Further, the NFL calendar lends itself to a three groupings of rest difference, with large spikes at +/-1 day, +/-3 days, and +/- 7 days difference. Alternatively, zero NFL games in our sample were played with a rest difference of 5 days.

Let $I(MNF)$, $I(Mini)$, and $I(Bye)$ be indicators for whether or not the home team $i$ (shown by $I(\cdot) = 1$) or visiting team $j$ (shown by $I(\cdot) = -1)$ have the rest advantage in a given game corresponding to Monday Night Football ($MNF$), mini-bye ($Mini$) and bye week ($Bye$), respectively. For example, if team $i$ has 10 days rest and team $j$ has 7 days rest, $I(MNF) = 0$, $I(Mini) = 1$, and $I(Bye) = 0$. Alternatively, if team $i$ has 7 days rest and team $j$ has 10 days rest, $I(MNF) = 0$, $I(Mini) = -1$, and $I(Bye) = 0$. Rest differential terms are assumed to be symmetric; e.g, the net benefit for a home team to a certain rest category is equivalent to had that same rest benefit been given to the visiting team. Additionally, rest terms are not mutually exclusive, such that if Team $i$ played the previous Monday (e.g, six days rest) and Team $j$ is off a bye, $I(MNF) = -1$ and $I(Bye) = -1$. The three indicator values for each combination of home and away rest days are also provided in Table \ref{tab:daysrest}. 

Model \ref{eq:model1} uses each of the three rest terms in modeling $Y_{s,ij}$ from 2002 to 2023, such that 
\begin{multline}
      E[Y_{s,ij}] = \theta_{s,i} - \theta_{s,j} + \mu_{HA, s} +
      \alpha_{MNF} \times I(MNF) + \alpha_{Mini} \times I(Mini) + \alpha_{Bye} \times I(Bye). \\
      \owntag[model1]{Constant Bye Effect, Point Differential} \end{multline}
\noindent In Model \ref{eq:model1}, $\alpha_{MNF}$, $\alpha_{Mini}$, and $\alpha_{Bye}$ reflect the point advantage of a team playing with each of the three rest advantages. 

Model \ref{eq:model2} splits the $Bye$ effect into two estimates, pre and post 2011 CBA change, such that
\begin{multline}
      E[Y_{s,ij}] = \theta_{s,i} - \theta_{s,j} + \mu_{HA, s}+ 
      \alpha_{MNF} \times I(MNF) +  \alpha_{Mini} \times I(Mini) + \\
      \alpha_{Bye, pre} \times I(Bye) \times I(S\leq2010) + \alpha_{Bye, post} \times I(Bye) \times I(S\geq2011). \\ \owntag[model2]{Pre/Post CBA Bye Effect, Point Differential}   
\end{multline}
\noindent In Model \ref{eq:model2},  $I(S\leq2010)$ and $I(S\geq2011)$ are indicators for pre and post 2011 CBA seasons. Finally, in Model \ref{eq:model2}, $\alpha_{Bye, pre}$ represents the point value of the bye advantage prior to the CBA change, and $\alpha_{Bye, post}$ reflects the point value of the bye week after the CBA change.  

\subsubsection{State Space models of betting odds}

Model's \ref{eq:model1} and \ref{eq:model2} use point differential as the outcome. In order to isolate the impact of rest on betting markets, we fit similar models using pre-game point spreads.

Let $Z_{s,ij}$ be the pre-game point spread between $i$ and $j$ in season s. As in Model's \ref{eq:model1} and \ref{eq:model2}, $i, j = 1, ... 32$, representing each of the NFL's current 32 franchises, and $s = 2002, ... 2023$. Our first model of the point spread, Model \ref{eq:model3}, assumes a constant $Bye$ effect in betting markets between 2002 and 2023, such that 

\begin{multline}
      E[Z_{s,ij}] = \theta_{s,i} - \theta_{s,j} +  \mu_{HA, s}  + 
      \alpha_{MNF} \times I(MNF) +  \alpha_{Mini} \times I(Mini) + \alpha_{Bye} \times I(Bye). \\
      \owntag[model3]{Constant Bye Effect, Point Spread}
\end{multline}

\noindent In Model \ref{eq:model3}, $\alpha_{MNF}$, $\alpha_{Mini}$, and $\alpha_{Bye}$ reflect the relative change in point spread given to teams with each of the three rest advantages. Akin to Model's \ref{eq:model1} and \ref{eq:model2}, $\theta_{s,i}$ and $\theta_{s,j}$ reflect season-level team strength (relative impact in point spread) for home team $i$ and away team $j$. Additionally, $\mu_{HA, s}$ reflects the season-level home advantage given to the home team in betting markets. 

Model \ref{eq:model4} splits the $Bye$ effect on the point spread before and after 2011, such that

\begin{multline}
      E[Z_{s,ij}] = \theta_{s,i} - \theta_{s,j} +  \mu_{HA, s}  + 
      \alpha_{MNF} \times I(MNF) + \alpha_{Mini} \times I(Mini) + \\
      \alpha_{Bye, pre} \times I(Bye) \times I(S\leq2010) + \alpha_{Bye, post} \times I(Bye) \times I(S\geq2011).\\ \owntag[model4]{Pre/Post CBA Bye Effect, Point Spread}   
\end{multline}

\noindent The coefficients in Model \ref{eq:model4} for the $Bye$ terms, $\alpha_{Bye, pre}$ and $\alpha_{Bye, post}$ reflect the relative impact on the point spread of the bye week rest advantage before and after the CBA change.

\subsubsection{Bayesian fits of state space models using Stan}

Each of Models \ref{eq:model1}-\ref{eq:model4} are fit using Stan, an open-source statistical software for Bayesian computing. There are several strengths to using a Bayesian approach. First, our primary interest lies in both the estimated rest effects and the probability these effects are greater (or less) than zero. Bayesian inference results in interpretable estimates alongside the direct calculations of these implied probabilities. Second, and related, is estimating the probability that $Bye$ week effect dropped after the 2011 CBA change. This can be directly calculated by using the posterior distributions of $\alpha_{Bye, pre}$ and $\alpha_{Bye, pre}$. Finally, a Bayesian paradigm is essential to use a state-space model with autoregressive priors as our model specification. This allows for dynamic estimates of team strength, thus preserving information in team ability from one year to the next. 

We use weakly informative prior distributions for all parameters of interest, and do not impose any outside knowledge in parameter estimation, such that

$$
\begin{aligned}
  \theta_{2002,i} &\sim N(0,\sigma_{\text{teamstrength}}^2) ~\forall~i~~~\text{(Team strengths in 2002)} \\
  \sigma_{\text{teamstrength}} &\sim \text{HalfNormal}(0,5^2)~~~\text{(Variance in team strength)}  \\
  \sigma_{game}  &\sim \text{HalfNormal}(0,5^2) ~~~\text{(Variance in point differential/point spread)} \\
  \gamma  &\sim \text{Uniform}(0,1) ~~~\text{(Autoregressive team strength parameter)} \\ \\
  \alpha_{HA\ Trend} &\sim N(0,5^2) ~~~\text{(Home advantage linear trend)} \\
  \alpha_{HA\ Intercept} &\sim N(0,5^2) ~~~\text{(Home advantage intercept)} \\ \\
  \alpha_{MNF} &\sim N(0,5^2) ~~~\text{(MNF rest advantage)} \\
  \alpha_{Mini} &\sim N(0,5^2) ~~~\text{(Mini rest advantage)} \\
  \alpha_{Bye} &\sim N(0,5^2) ~~~\text{(Bye rest advantage)} \\
  \alpha_{Bye, pre} &\sim N(0,5^2) ~~~\text{(Bye rest advantage, pre CBA bye effect)} \\
  \alpha_{Bye, post} &\sim N(0,5^2) ~~~\text{(Bye rest advantage, post CBA bye effect)} \\
\end{aligned}
$$

Posterior distributions of each parameter are estimated using Markov Chain Monte Carlo (MCMC). We used 4 parallel chains of 3000 iterations, with a burn in of 1000 draws. For model convergence, we visually examine trace plots (e.g, time series plots of the Markov Chains). 

All data and code for both (i) the data cleaning/filtering and (ii) the statistical analysis is presented on Github at \url{https://github.com/ThompsonJamesBliss/rest-advantage-in-amer-football}. 

\subsubsection{Model comparison}

Our primary points of comparison for each of the two pairs of models (e.g, Model \ref{eq:model1} to \ref{eq:model2} and Model \ref{eq:model3} to \ref{eq:model4}) are the posterior draws of $\alpha_{Bye, pre}$ and $\alpha_{Bye, post}$. For example, $Pr(\alpha_{Bye, post} > \alpha_{Bye, pre})$ corresponds to the probability that the bye advantage increased. Likewise, $Pr(\alpha_{Bye, post} < \alpha_{Bye, pre})$ corresponds to the probability that the bye advantage decreased. 

Additionally, and as recommended by \cite{vehtari2017practical}, we use log pointwise predictive density (ELPD), which is estimated via leave-one-out cross validation, in order to more formally compare Model \ref{eq:model1} to Model \ref{eq:model2} and Model \ref{eq:model3} to Model \ref{eq:model4}. This is done by computing the log likelihood of each outcome (either the point differential or the point spread) using each posterior sample. One benefit of using ELPD for model comparison purposes is that standard error estimates are readily attainable, which allows for differences in model fit statistics to be standardized.

\section{Results} \label{sec:3}

\subsection{Data overview}  

There were 5679 regular season NFL games from 2002 to 2023 (256 games per season from 2002 to 2020, 272 per season in 2021 and 2023, and 271 in 2022). Game information (date, home team, away team, final score) was obtained using internal NFL data sources, and point spread information stems from nflreadR, which generates lines from an aggregate of sports books \citep{nflreadR}. All data are found in our Github repository at \url{https://github.com/ThompsonJamesBliss/rest-advantage-in-amer-football}. 

The most common rest advantage was the $MNF$ benefit, allocated to 410 away teams and 290 home teams from 2002 to 2023 (compared to a total of 526 $Mini$ and 593 $Bye$ advantages during this time period). Due to past scheduling practices, roughly 80 percent of $Mini$ bye rest edges came after 2011.\footnote{No more than three Thursday games were played each season from 2002 to 2005 and no more than 10 Thursday games were played from 2006 to 2010. From 2012 onwards, the Thursday Night NFL package has expanded to 16 to 18 Thursday games per season.}

A total of 52 games (0.9\% percent) were dropped for having rest counts that failed to fit in our categories. As referenced in Section \ref{sec:3},  most of these games were rescheduled games or games following rescheduled games due to the COVID-19 pandemic. Altogether, 46 of the 52 dropped games were played in either 2020 or 2021. The complete list of dropped games is shown in Table \ref{tab:gamesdropped} in the Appendix. 

Table \ref{tab:statsummary} shows the counts, raw point differential, win percentage, expected win percentage, and cover percentage for each of our four rest categories, split by side (home/away) and era (2002-2010, 2011-2023, and overall). Expected win percentages are calculated by converting betting market money lines for each team into a percent (see \cite{lopez2018} for an example), and reflect the percent of times each team is expected to win. Given differences in team strength in various NFL game windows, expected win percentage can be used as a reference point for comparing to actual win percentage. We count ties as half a win for both game win-loss and cover win-loss outcomes.  

\begin{center}
    (Table \ref{tab:statsummary} here)
\end{center} 

As baselines, home teams with equivalent rest in Table \ref{tab:statsummary} had an average point differential of +2.69 from 2002 to 2010 and +1.83 from 2011 to 2023. Alternatively, home teams with a $Bye$ advantage had an average point differential of +4.45 from 2002 to 2010 and +1.76 from 2011 to 2023. Without accounting for team strength, this suggests a decline in $Bye$ benefit from +1.76 (e.g, +4.45 to +2.69) points per game to -0.07 (+1.76 to +1.83) points per game for the home team. 

Both home and away teams with a $Bye$ advantage covered more often than expected from 2002 to 2010 (55.8\% for the home teams with a bye advantage, 56.9\% for road teams with a bye advantage). This is in line with the betting market inefficiency suggested by \cite{sung2014national}. In recent seasons, that edge has flipped; home teams with a $Bye$ edge covered only 44.6\% of the time between 2011 and 2023 (away teams 52.7\%).  

\subsection{Model checks and validation}

Trace plots for each of our parameters of interest, done for both the point differential and point spread models, are shown in the Appendix in Figures \ref{fig:model_1_trace} to \ref{fig:model_4_trace}. There is no evidence of a lack of convergence with any of the parameters in any of our four models.

\begin{figure}[hbt!] 
    \centering
    \includegraphics[width=\textwidth]{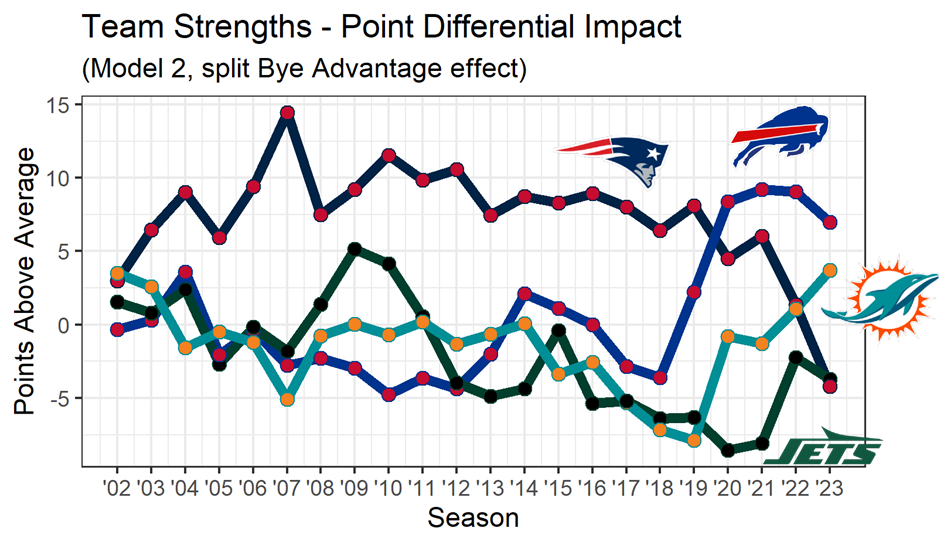}
    \caption{Team strengths for AFC East by season, estimated using our state space model of point differential. Results roughly match intuition, with the New England Patriots as the AFC East's best team for the majority of two decades, only to be replaced by the Buffalo Bills from 2020 to 2023.}.
    \label{fig:team_strength}
\end{figure}

As an additional sensibility check, Figure \ref{fig:team_strength} shows the average team strength in each season for each of the four teams in the AFC East, the New England Patriots, Buffalo Bills, Miami Dolphins, and New York Jets, as generated using Model \ref{eq:model2}. Throughout the 2000s and most of the 2010s, the Patriots finished first in this division in points above average (according to our Model \ref{eq:model2}), peaking in the 2007 season (14.5 points better than an average team). This reflects results on the field, as the actual Patriots won the AFC East 15 seasons during these years. From 2020 to 2023, the Bills had the highest points above average according to Model \ref{eq:model2}. Again, this reflects reality, as the actual Bills won the division during each of these four seasons. Though this does not formally validate our team strength estimates, the resemblance to known team ability is reassuring, allowing us to more comfortably estimate rest advantage values. 
\begin{center}
    (Figure \ref{fig:team_strength} here)
\end{center}

\subsection{Model comparison}

In the posterior draws from Model \ref{eq:model2}, $Pr(\alpha_{Bye, post} < \alpha_{Bye, pre}) = 0.966$. In other words, there is a 96.6\% chance that the $Bye$ advantage in point differential declined after the 2011 CBA change. Alternatively, using Model \ref{eq:model4}, $Pr(\alpha_{Bye, post} > \alpha_{Bye, pre}) = 0.988$, suggesting a near 99\% probability that betting markets $increased$ their valuation of the point spread. 

Table \ref{tab:loosummary} compares each of the two pairs of models using ELPD, calculated using leave-one-out cross validation and the posterior distributions of each model. 

\begin{center}
    (Table \ref{tab:loosummary} here)
\end{center}

Differences in ELPD are approximately normal; using this criterion, Model \ref{eq:model2} is expected to have better predictive performance than Model \ref{eq:model1} (0.14 standard errors away from 0). Alternatively, Model \ref{eq:model4} is expected to have better predictive performance than Model \ref{eq:model3} (1.27 standard errors away from 0). 

Overall, model comparison tools suggest that, for both point differential and point spread outcomes, 2011-2023 reflects differing impacts of the $Bye$ advantage. 

\subsection{Estimated impact of rest}

\begin{figure}[hbt!] 
    \centering
    \includegraphics[width=\textwidth]{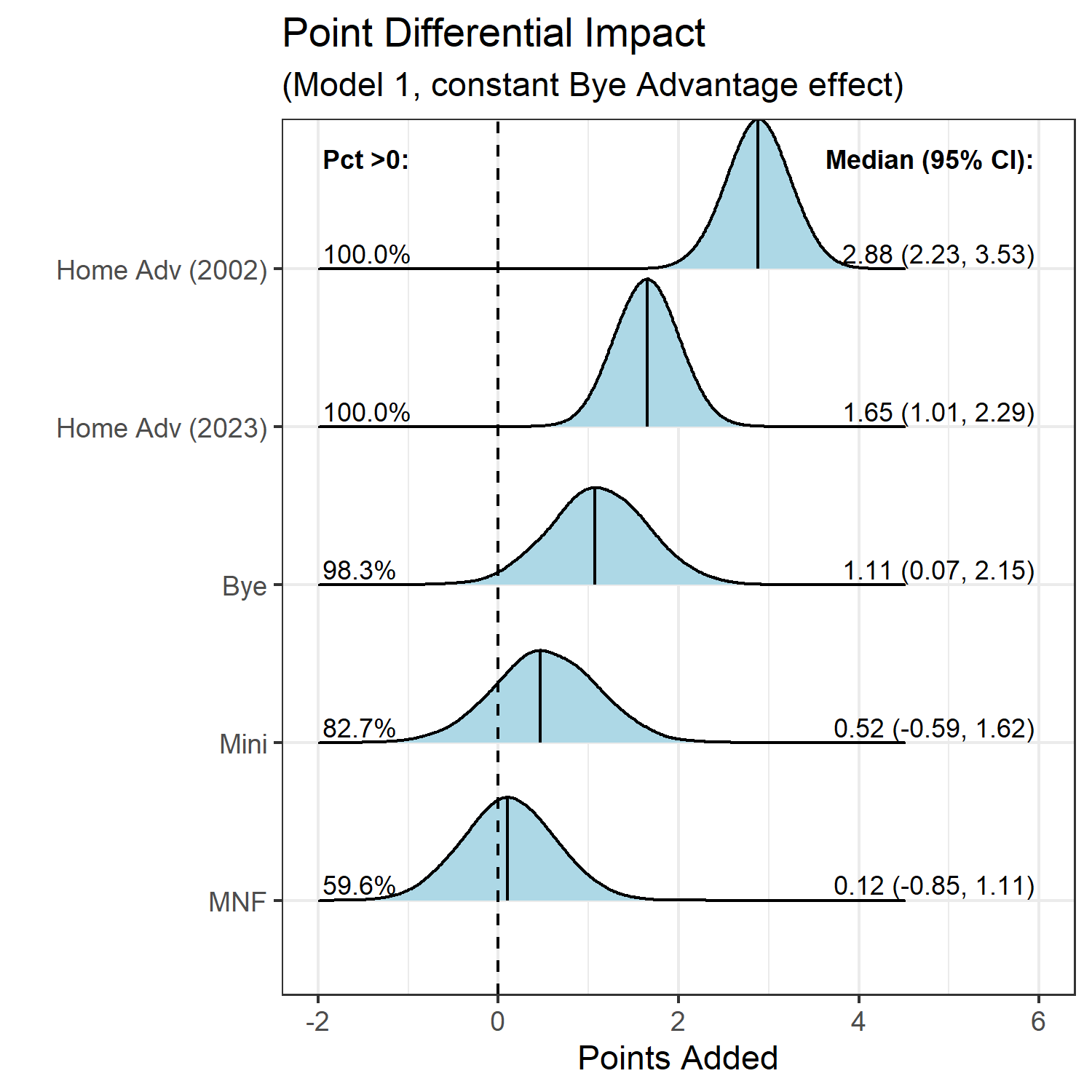}
    \caption{Distributions of posterior draws for each parameter of interest in the constant $Bye$ advantage model fit on point differential. For each parameter, the figure is labelled with (i) the proportion of draws greater than 0 and (ii) the median and 95\% credible intervals.}
    \label{fig:model_1}
\end{figure}

\begin{figure}[hbt!] 
    \centering
    \includegraphics[width=\textwidth]{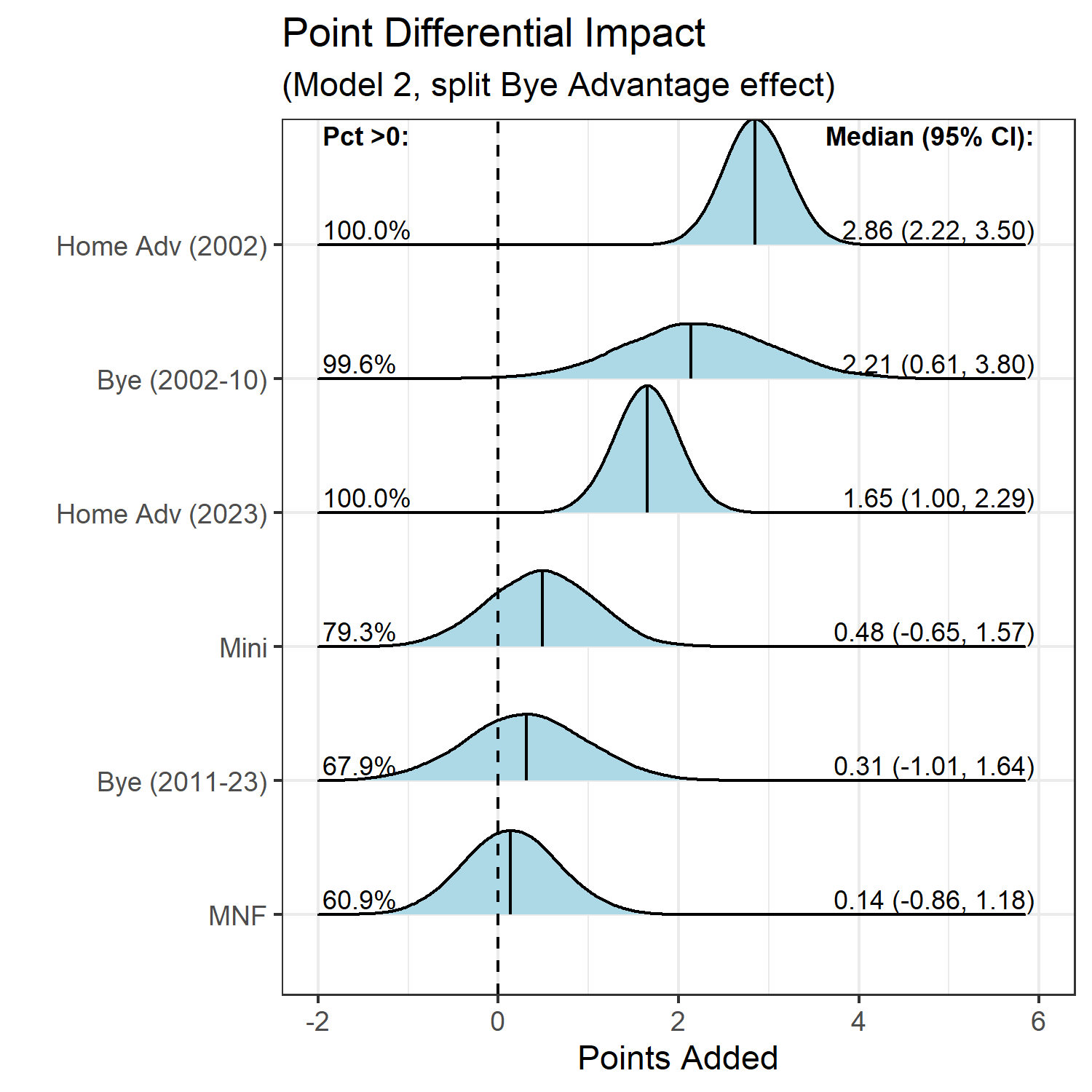}
    \caption{Distributions of posterior draws for each parameter of interest in the split $Bye$ advantage model (2002-2010, 2011-2023) fit on point differential. For each parameter, the figure is labelled with (i) the proportion of draws greater than 0 and (ii) the median and 95\% credible intervals.}.
    \label{fig:model_2}
\end{figure}

\begin{figure}[hbt!] 
    \centering
    \includegraphics[width=\textwidth]{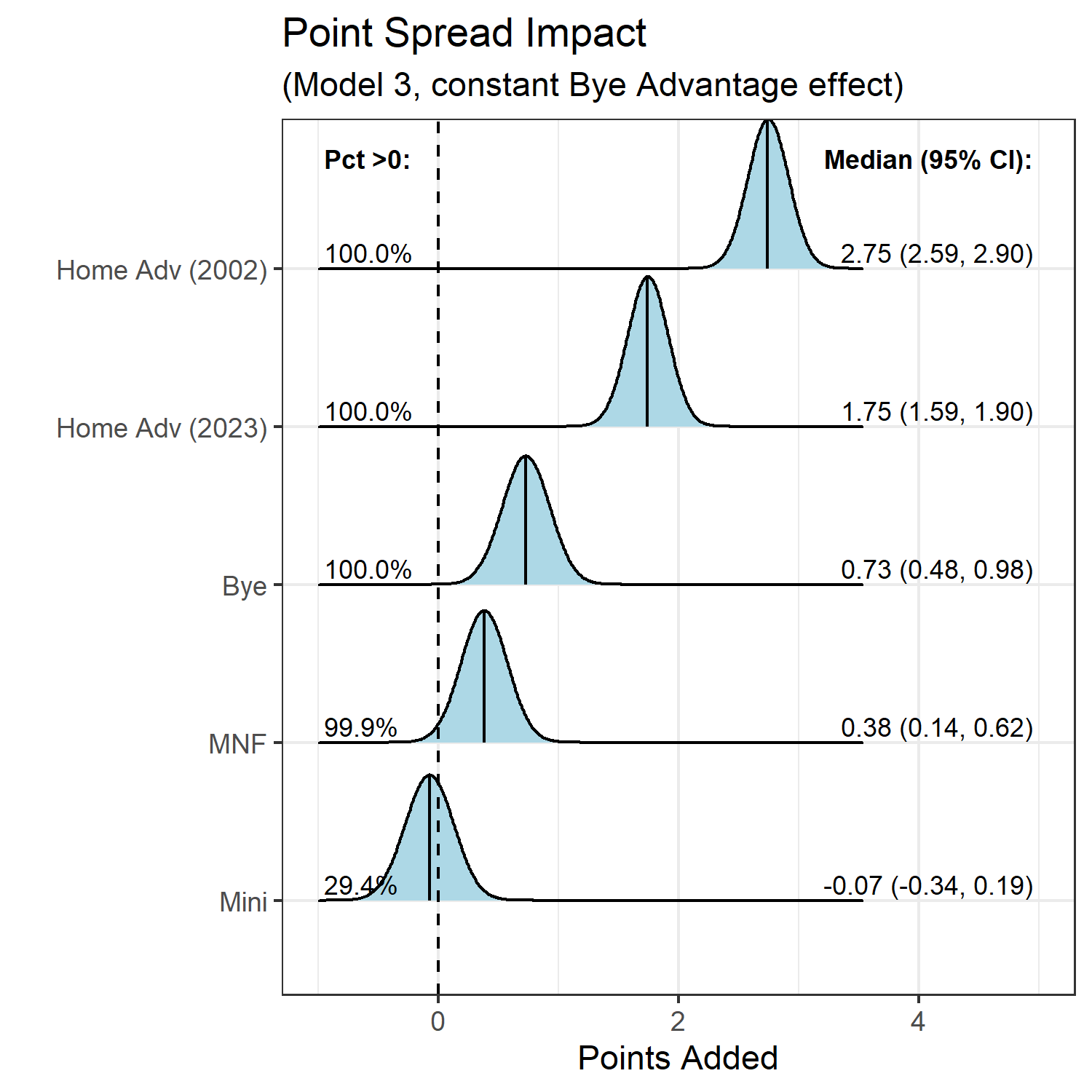}
    \caption{Distributions of posterior draws for each parameter of interest in the constant $Bye$ advantage model fit on the point spread. For each parameter, the figure is labelled with (i) the proportion of draws greater than 0 and (ii) the median and 95\% credible intervals.}.
    \label{fig:model_3}
\end{figure}

\begin{figure}[hbt!] 
    \centering
    \includegraphics[width=\textwidth]{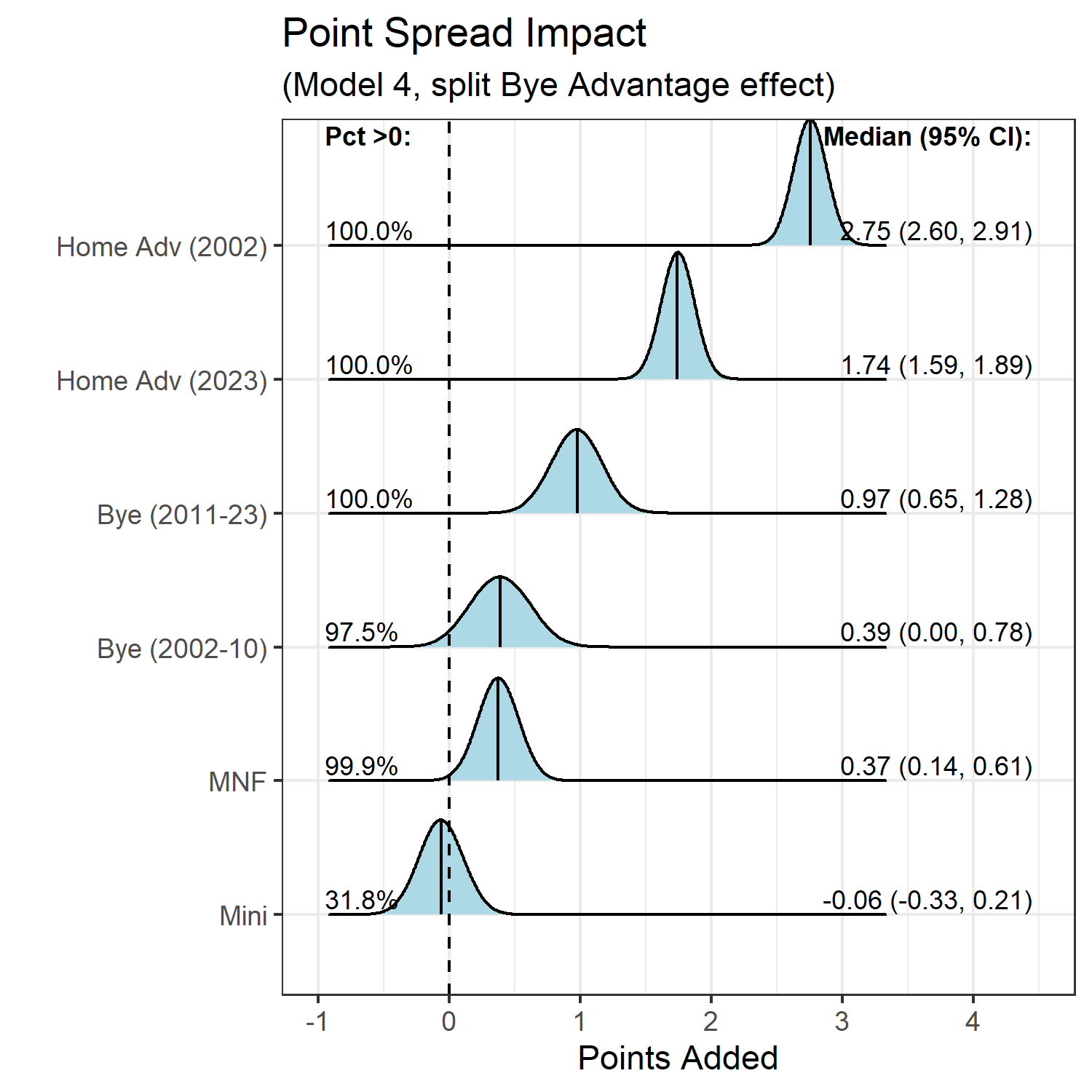}
    \caption{Distributions of posterior draws for each parameter of interest in the split $Bye$ advantage model (2002-2010, 2011-2023) fit on the point spread. For each parameter, the figure is labelled with (i) the proportion of draws greater than 0 and (ii) the median and 95\% credible intervals.}
    \label{fig:model_4}
\end{figure}

Figures \ref{fig:model_1} to \ref{fig:model_4} show the distributions of posterior draws for each parameter of interest from Models \ref{eq:model1} to \ref{eq:model4}. Each figure is labelled with (i) the proportion of draws greater than 0, and (ii) the median posterior draw and 95\% credible interval for each parameter. The median reflects our best estimate of each rest advantage, the credible intervals correspond to a reasonable set of bounds, and the proportions reflect probabilities that each rest category offers a benefit. As context for the relative impact of the rest advantages, Figures \ref{fig:model_1} to \ref{fig:model_4} also show the posterior draws for $\mu_{HA, 2002}$ and $\mu_{HA, 2023}$, the estimated home advantages in 2002 and 2023, respectively. 
\begin{center}
    (Figures \ref{fig:model_1} to \ref{fig:model_4} here)
\end{center}

\subsubsection{$Bye$ Advantage}

Model results suggest that the $Bye$ advantage provided a significant edge before the CBA change, but that most, if not all, of the rest benefit has since been eliminated. Using Model \ref{eq:model2}, prior to 2011, there is a 99.6\% probability that the $Bye$ advantage had a positive effect on point differential, and our best estimate of the benefit is +2.21 points per game (95\% credible interval, 0.61 to 3.80). The effect size of this $Bye$ advantage is similar to that of the home advantage; that is, prior to 2011, it was as beneficial to play off a bye week as it was to play at home. 

Post 2011, the $Bye$ advantage does not provide a statistically detectable benefit. Model \ref{eq:model2} implies a 67.9\% probability that the $Bye$ advantage has a positive effect on point differential, with a median estimate of +0.31 points per game (95\% credible interval, -1.01 to 1.64). 

While the benefit of the $Bye$ edge has dropped post 2011, betting markets have moved in the opposite direction. Using Model \ref{eq:model4}, betting markets increased the value of the $Bye$ advantage from +0.39 points per game (95\% credible interval, 0.00 to 0.78) pre CBA to +0.97 points per game (95\% credible interval, 0.65 to 1.28) post CBA. 

\begin{figure}[hbt!] 
    \centering
    \includegraphics[width=\textwidth]{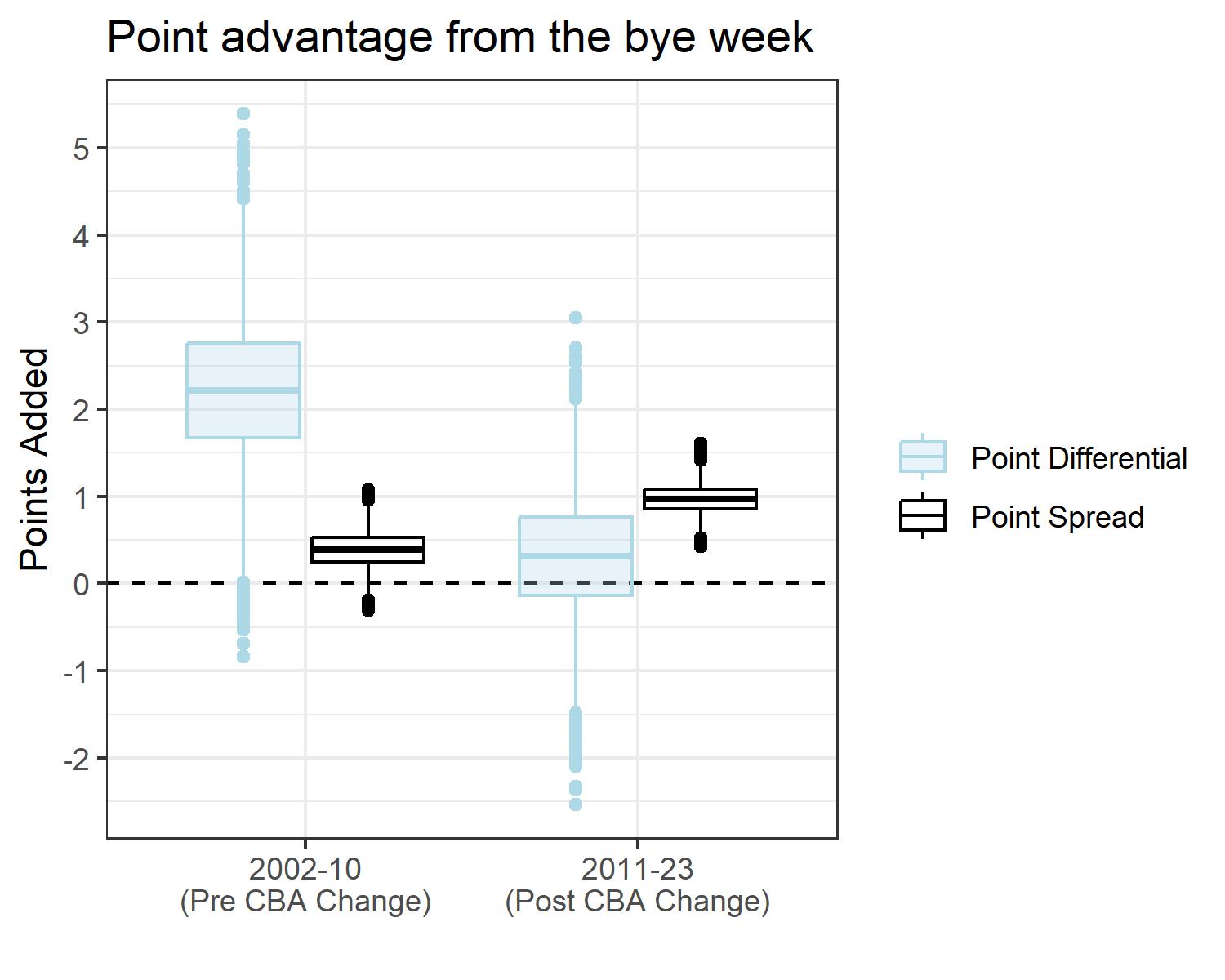}
    \caption{Boxplots showing the estimated impact of the $Bye$ advantage on both point differential (in shaded blue) and point spread (black). Benefits are split pre and post 2011, reflecting the shift in the NFL's Collective Bargaining Agreement on bye week rules. Pre-2011, betting markets undervalued the $Bye$ advantage by nearly two points per game, while post-2011, betting markets have overvalued the $Bye$ advantage by roughly half a point. Posterior draws are pulled from Model's \ref{eq:model2} and \ref{eq:model4}.}.
    \label{fig:boxplot_bye}
\end{figure}

Figure \ref{fig:boxplot_bye} shows boxplots of distributions to directly compare the change in bye week valuation pre and post 2011 CBA. The impact in point differential (and the drop in impact) is shown in blue, with the median decline in point differential value of nearly 1.9 points per game. In black is the point spread impact, which, over the same period, increased by roughly 0.6 points per game. 

\begin{center}
    (Figure \ref{fig:boxplot_bye} here)
\end{center}

The current valuation of +0.97 points per game for a $Bye$ advantage is notable, as it is in line with the estimated edge 
in point differential from Model \ref{eq:model1} of +1.11 points per game (95\% credible interval, 0.07 to 2.15). That is, the current betting market adjustment awarded to teams off a bye week directly corresponds to the $Bye$ effect on point differential in a statistical model fit over the past two decades. 

\subsubsection{$Mini$ and $MNF$ Advantages}

The $Mini$ bye advantage estimate from Model \ref{eq:model2} is +0.48 points per game (95\% credible interval, -0.65 to 1.57) in the point differential model, suggesting a small but non-significant benefit.  However, betting markets do not consider the $Mini$ bye to offer any advantage; the median posterior draw for the $Mini$ effect in Model \ref{eq:model4} is -0.06 (95\% credible interval, -0.33 to 0.21). In Model \ref{eq:model4}, there is only a 31.8\% probability the $Mini$ bye has a positive impact on the point spread.

Using the point differential model, the median posterior draw for the $MNF$ benefit is +0.14 points per game (95\% credible interval, -0.86 to 1.18). Though that estimated effect is not statistically significant, betting markets believe a $MNF$ rest benefit exists. Using Model \ref{eq:model4}, there is a 99.9\% chance the MNF edge has a significant impact on betting lines, with a best estimate of +0.37 points per game (95\% credible interval, 0.14 to 0.61). 

\section{Discussion} \label{sec:4}

The 2011 Collective Bargaining Agreement changed bye week regulations in the National Football League. Prior to the agreement, teams could practice and hold other organized activities; after the 2011 season, however, regular season bye weeks came with a guarantee of at least four days off. In place of practice, coaches and personnel often `self-scout,' analyzing their own tendencies in place of preparing for upcoming opponents \citep{selfscout}. Players, with a week off, travel or spend time outside the team facility with their families. 

The change in bye week policy provides a natural experiment to test the impact the the $Bye$ advantage in NFL game outcomes. We find that the majority of benefit to the bye week drops after 2011, from +2.21 points per game to +0.31 points per game. The advantage, which at one point was not picked up on by betting markets (found in both our results and in \cite{sung2014national}), no longer appears to provide clubs a significant competitive edge. Ironically, betting market numbers increased their valuation of the $Bye$ advantage impact, from +0.39 to +0.97 points per game, in this same time frame. 

This result has direct implications for clubs and the NFL schedule. If the primary benefit of the bye week was rest and recovery, we would have expected that $Bye$ effect to hold, as the rest and recovery allocated to players was unchanged after 2011. Instead, the drop in $Bye$ effect suggests that the majority of benefit to the bye was extra practice and preparation.

As a follow-up to our bye week results, we note that the 2011 CBA change only impacted regular season bye weeks. In postseason play, NFL teams with a bye edge still get an extra week to practice and prepare for an upcoming opponent. Though too small a sample of games to use in our modeling framework, teams with a postseason bye edge still appear to get a significant benefit. From 2002 to 2010, the 36 teams with a postseason bye went 22-10 (win pct, 0.611) with an average point differential of +5.17, while finishing 15-21 against the point spread (cover pct, 0.417). From 2011 to 2023, the 44 teams with a postseason bye went 34-10 (win pct, 0.773) with an average point differential of +7.34, finishing 21-22-1 against the point spread (cover pct, .489). Altogether, a postseason bye advantage appears to have remained consistent, if not grown, after the 2011 CBA. 

Results are mixed on the other two edges, the $Mini$ bye and the $MNF$ edge. Betting markets think it is equally likely that the $Mini$ bye advantage is actually a disadvantage, while the point differential model suggests the $Mini$ edge provides a non-significant benefit of +0.48 points per game. If a $Mini$ edge was indeed non-zero, there is reason to believe that it too has been mitigated. The 2020 NFL Collective Bargaining Agreement \citep{nflcba_2020} limited what teams could do after Thursday Night Football games, guaranteeing at least three days off (unless teams are playing on two consecutive Thursdays). In other words, the practice limitations that negatively impacted the $Bye$ week edge now apply to the $Mini$ advantage. As a result, it appears unlikely that any positive effect to the $Mini$ bye will exist moving forward.  

Betting market models imply roughly a third of a point benefit to the $MNF$ edge. In practice, our point differential model is less certain that this edge exists (+0.18 points per game). If the implication of our bye week findings is that practice and preparation is the principal driver of a competitive benefit, it stands that the $MNF$ advantage -- where one team gets a full week of practice, while the other receives at least one less day -- could be beneficial. However, relative to other NFL competitive advantages (such as playing at home, or the $Bye$ edge before 2011), any $MNF$ benefit is orders of magnitude smaller. 

While estimating the home advantage was not a primary goal of our research, in accounting for home advantage in our models, we note a few interesting results. First, while betting markets were behind trends in the $Bye$ advantage impact, for the home advantage, betting markets appear to be near perfect proxies for point differential results. For example, Model \ref{eq:model2} suggests that the average benefit to a 2023 home game was +1.65 points per game; Model \ref{eq:model4} finds that betting markets adjusted the 2023 point spread an estimated 1.74 points per game. Additionally, we note that each of Models \ref{eq:model1} to \ref{eq:model4} suggest a decline of roughly a point per game in the home advantage. This mirrors the results of \cite{benz2024comprehensive}, who found a similar decline. 

When comparing these findings to other sports, the competitive impact of back to back games in the NBA and NHL stands out as notable. If a difference of 7 and 14 days in the NFL is not significant for rest and recovery, it stands to reason that there's some threshold -- two days? three days? -- that it takes NFL bodies to recover and practice in order to reach a similar competitive performance. Related, studies have repeatedly shown that playing on Thursday Night Football (4 days between games) does not increase injury risk (see \cite{baker2019thursday} and \cite{binney2020incorrect} follow-up, and \cite{perez2020effect}) or concussion risk \citep{teramoto2017game},

One close corollary is professional soccer, as European football policy is increasingly subject to unique, broadcast-driven schedules. In the English Premier League, for example, the percentage of games played on Saturday's has dropped from 54.2 percent of the schedule (2019) to 51.3 percent (2024). In place, 11 games in 2024 were played on Thursday (2 in 2019) and 102 on Sunday (86 in 2019). More games on unique days entails an increase in the dispersion in relative rest. Additionally, Premier League teams received a `winter break` in the 2023-2024 season. During this period, half the teams played on one weekend, and half the teams played the next weekend. Given our findings on the NFL's bye week, it seems unlikely that this winter break would provide a competitive benefit if players were given the time off.

\section*{Conflict of Interest Statement}

Both authors are full time employees of the National Football League. Results of this paper have been shared with league decision makers as they look to improve the schedule-making process. We have shared our data and code with the community in the hopes that others can build and/or expand our findings. 

\section*{Author Contributions}
ML: Writing - original draft, review and editing, Conceptualization, Methodology, Project Administration, Validation; TB: Writing - review and editing, Software, Visualization

\section*{Acknowledgments}
Thank you to our partners in Broadcast and Data and Analytics at the NFL League Office for the origination of these ideas, as well as suggestions on definitions and an analysis framework. Additionally, thanks to several members of the football analytics community for their feedback on original drafts.

\section*{Data Availability Statement}
The datasets analyzed for this study (as well as the code for all models and figures) can be found on Github at \url{https://github.com/ThompsonJamesBliss/rest-advantage-in-amer-football}.

\bibliographystyle{Frontiers-Harvard} 

\bibliography{test}


\clearpage

\section*{Tables}

\begin{table}[hbt!] 
\centering
\begin{tabular}{rrrrrrr}
  \hline
Home rest (days) & Away rest (days) & Difference (days) & $Bye$ & $Mini$ & $MNF$ & Num. games \\ 
  \hline
  4 & 4 & 0& 0 & 0 & 0 & 243 \\ 
  5 & 5 & 0& 0 & 0 & 0 & 6 \\ 
  6 & 6 & 0& 0 & 0 & 0 & 86 \\ 
  7 & 7 & 0& 0 & 0 & 0 & 2839 \\ 
  8 & 8 & 0& 0 & 0 & 0 & 250 \\ 
  10 & 10 & 0& 0 & 0 & 0 & 3 \\ 
  11 & 11 & 0& 0 & 0 & 0 & 1 \\ 
  14 & 14 & 0& 0 & 0 & 0 & 35 \\ 
  15 & 15 & 0& 0 & 0 & 0 & 4 \\ 
  5 & 6 & -1& 0 & 0 & -1 & 2 \\ 
  6 & 5 & 1& 0 & 0 & 1 & 2 \\ 
  6 & 7 & -1& 0 & 0 & -1 & 340 \\ 
  7 & 6 & 1& 0 & 0 & 1 & 250 \\ 
  7 & 8 & -1& 0 & 0 & 0 & 44 \\ 
  8 & 7 & 1& 0 & 0 & 0 & 47 \\ 
  8 & 9 & -1& 0 & 0 & 0 & 6 \\ 
  9 & 8 & 1& 0 & 0 & 0 & 2 \\ 
  10 & 9 & 1& 0 & 0 & 0 & 1 \\ 
  13 & 12 & 1& 0 & 0 & 0 & 1 \\ 
  13 & 14 & -1& 0 & 0 & 0 & 6 \\ 
  14 & 13 & 1& 0 & 0 & 0 & 3 \\ 
  6 & 8 & -2& 0 & 0 & -1 & 10 \\ 
  7 & 9 & -2& 0 & -1 & 0 & 6 \\ 
  8 & 6 & 2& 0 & 0 & 1 & 3 \\ 
  8 & 10 & -2& 0 & -1 & 0 & 6 \\ 
  9 & 7 & 2& 0 & 1 & 0 & 2 \\ 
  10 & 8 & 2& 0 & 1 & 0 & 2 \\ 
  6 & 9 & -3& 0 & -1 & -1 & 5 \\ 
  7 & 10 & -3& 0 & -1 & 0 & 217 \\ 
  8 & 11 & -3& 0 & -1 & 0 & 33 \\ 
  9 & 6 & 3& 0 & 1 & 1 & 6 \\ 
  10 & 7 & 3& 0 & 1 & 0 & 164 \\ 
  10 & 13 & -3& -1 & 1 & 0 & 4 \\ 
  11 & 8 & 3& 0 & 1 & 0 & 21 \\ 
  13 & 10 & 3& 1 & -1 & 0 & 2 \\ 
  6 & 10 & -4& 0 & -1 & -1 & 23 \\ 
  7 & 11 & -4& 0 & -1 & 0 & 1 \\ 
  10 & 6 & 4& 0 & 1 & 1 & 14 \\ 
  10 & 14 & -4& -1 & 1 & 0 & 6 \\ 
  11 & 7 & 4& 0 & 1 & 0 & 1 \\ 
  11 & 15 & -4& -1 & 1 & 0 & 4 \\ 
  14 & 10 & 4& 1 & -1 & 0 & 7 \\ 
  15 & 11 & 4& 1 & -1 & 0 & 3 \\ 
  7 & 13 & -6& -1 & 0 & 0 & 22 \\ 
  13 & 7 & 6& 1 & 0 & 0 & 22 \\ 
  6 & 13 & -7& -1 & 0 & -1 & 4 \\ 
  7 & 14 & -7& -1 & 0 & 0 & 206 \\ 
  8 & 15 & -7& -1 & 0 & 0 & 24 \\ 
  13 & 6 & 7& 1 & 0 & 1 & 1 \\ 
  14 & 7 & 7& 1 & 0 & 0 & 223 \\ 
  15 & 8 & 7& 1 & 0 & 0 & 25 \\ 
  6 & 14 & -8& -1 & 0 & -1 & 26 \\ 
  14 & 6 & 8& 1 & 0 & 1 & 14 \\ 
   \hline
\end{tabular}
\caption{Counts of days rest for each of the home and away teams for each NFL game in our sample. Also provided are which, if any, rest advantage category those games fit into ($Bye$, $Mini$, and $MNF$). Indicators of 1 reflect the rest edge going to the home team, while indicators of -1 correspond to an away team benefit.}
\label{tab:daysrest}
\end{table}

\begin{table}[hbt!] 
\centering
\begin{tabular}{lllrrrrr}
  \hline
Side & Type & era & n & Point Diff & Win Pct & Exp Win Pct & Cover \% \\ 
  \hline
 \multirow{12}{*}{Away} &  & 2002-10 & 1671 & -2.689 & 0.423 & 0.432 & 0.511 \\ 
      & Equivalent Rest & 2011-23 & 2261 & -1.826 & 0.447 & 0.449 & 0.513 \\ 
      &  & All & 3932 & -2.193 & 0.437 & 0.442 & 0.512 \\ \cline{2-8}
      &  & 2002-10 &  158 & -4.380 & 0.405 & 0.380 & 0.547 \\ 
      & MNF Rest & 2011-23 &  252 & -3.333 & 0.421 & 0.439 & 0.498 \\ 
      &  & All &  410 & -3.737 & 0.415 & 0.416 & 0.517 \\ \cline{2-8}
      &  & 2002-10 &   70 & 0.857 & 0.529 & 0.461 & 0.500 \\ 
      & Mini (TNF) Rest & 2011-23 &  233 & -2.082 & 0.431 & 0.451 & 0.509 \\ 
      &  & All &  303 & -1.403 & 0.454 & 0.454 & 0.507 \\ \cline{2-8}
      &  & 2002-10 &  109 & -0.248 & 0.459 & 0.441 & 0.569 \\ 
      & Bye Rest & 2011-23 &  187 & -1.684 & 0.473 & 0.456 & 0.527 \\ 
      &  & All &  296 & -1.155 & 0.468 & 0.450 & 0.542 \\ \hline
 \multirow{12}{*}{Home}      &  & 2002-10 & 1671 & 2.689 & 0.577 & 0.568 & 0.489 \\ 
      & Equivalent Rest & 2011-23 & 2261 & 1.826 & 0.553 & 0.551 & 0.487 \\ 
      &  & All & 3932 & 2.193 & 0.563 & 0.558 & 0.488 \\ \cline{2-8}
      &  & 2002-10 &  126 & -0.349 & 0.540 & 0.528 & 0.472 \\ 
      & MNF Rest & 2011-23 &  164 & 3.341 & 0.561 & 0.559 & 0.518 \\ 
      &  & All &  290 & 1.738 & 0.552 & 0.545 & 0.498 \\ \cline{2-8}
      &  & 2002-10 &   42 & 1.048 & 0.452 & 0.546 & 0.488 \\ 
      & Mini (TNF) Rest & 2011-23 &  181 & 3.840 & 0.561 & 0.578 & 0.514 \\ 
      &  & All &  223 & 3.314 & 0.540 & 0.572 & 0.509 \\ \cline{2-8}
      &  & 2002-10 &  139 & 4.453 & 0.651 & 0.576 & 0.558 \\ 
      & Bye Rest & 2011-23 &  158 & 1.759 & 0.589 & 0.588 & 0.446 \\ 
      &  & All &  297 & 3.020 & 0.618 & 0.582 & 0.498 \\ 
   \hline
\end{tabular}
\caption{Summary statistics showing home and away team performance (avg point differential, win percentage, expected win percentage, and cover percentage) from 2002 to 2023. Rest advantages are split into pre and post 2011 Collective Bargaining Agreement (CBA), which changed the rules on when teams could practice during bye weeks. Expected win percentage is calculated using sportsbook betting odds.}
\label{tab:statsummary}
\end{table}

\begin{table}[hbt!] 
\centering
\begin{tabular}{lccrrrr}
  \hline
Outcome &  Model & ELPD (Difference*) & SE (Difference*) & \# SE\\ 
  \hline
  \multirow{ 2}{*}{Point Differential}  & Model 2 & 0 & 0 & 0  \\ 
           & Model 1 & -0.26 & 1.80 & 0.14 \\ 
  \hline
  \multirow{ 2}{*}{Point Spread}   & Model 4 & 0 & 0 & 0 \\ 
   & Model 3 & -2.54 & 2.00 & 1.27 \\ 
  \hline
   \hline
\end{tabular}
\caption{Model comparison table using log pointwise predictive density (ELPD), as estimated using leave-one-out cross validation. The better scoring models reflect those with higher log likelihoods, with the differences between the two models (*) assessed using standard errors. This metric suggests Model's 2 and 4 (with a pre and post $Bye$ week effect) are better predictive models for point differential and point spread than Model's 1 and 3 (constant $Bye$ week effect), respectively.}
\label{tab:loosummary}
\end{table}

\clearpage

\section*{Figures}

\clearpage

\section*{Appendix} 

\begin{figure}[hbt!] 
    \centering
    \includegraphics[width=\textwidth]{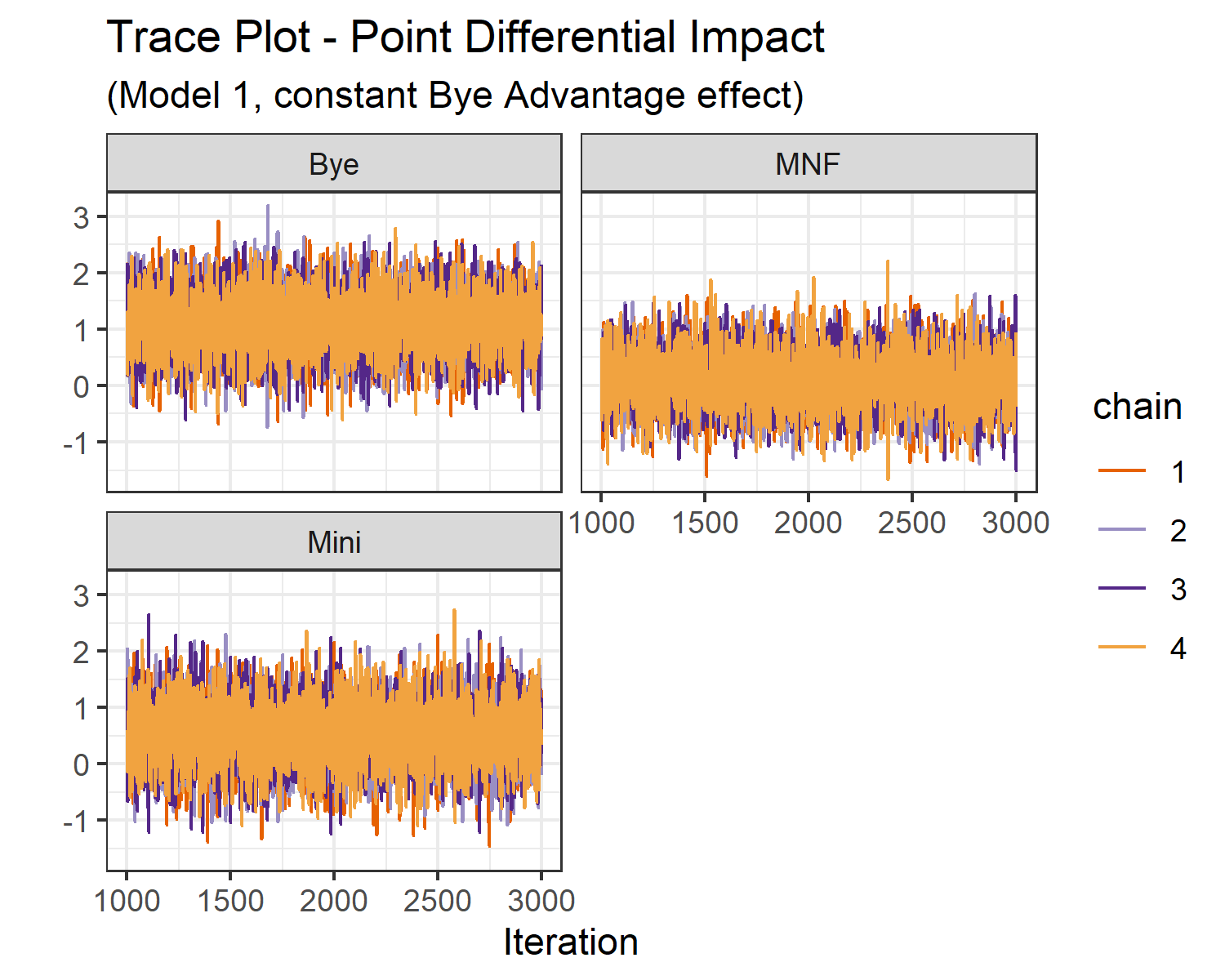}
    \caption{Trace plot for each rest parameter of interest for Model \ref{eq:model1}. There is no evidence of a lack of convergence.}.
    \label{fig:model_1_trace}
\end{figure}

\begin{figure}[hbt!] 
    \centering
    \includegraphics[width=\textwidth]{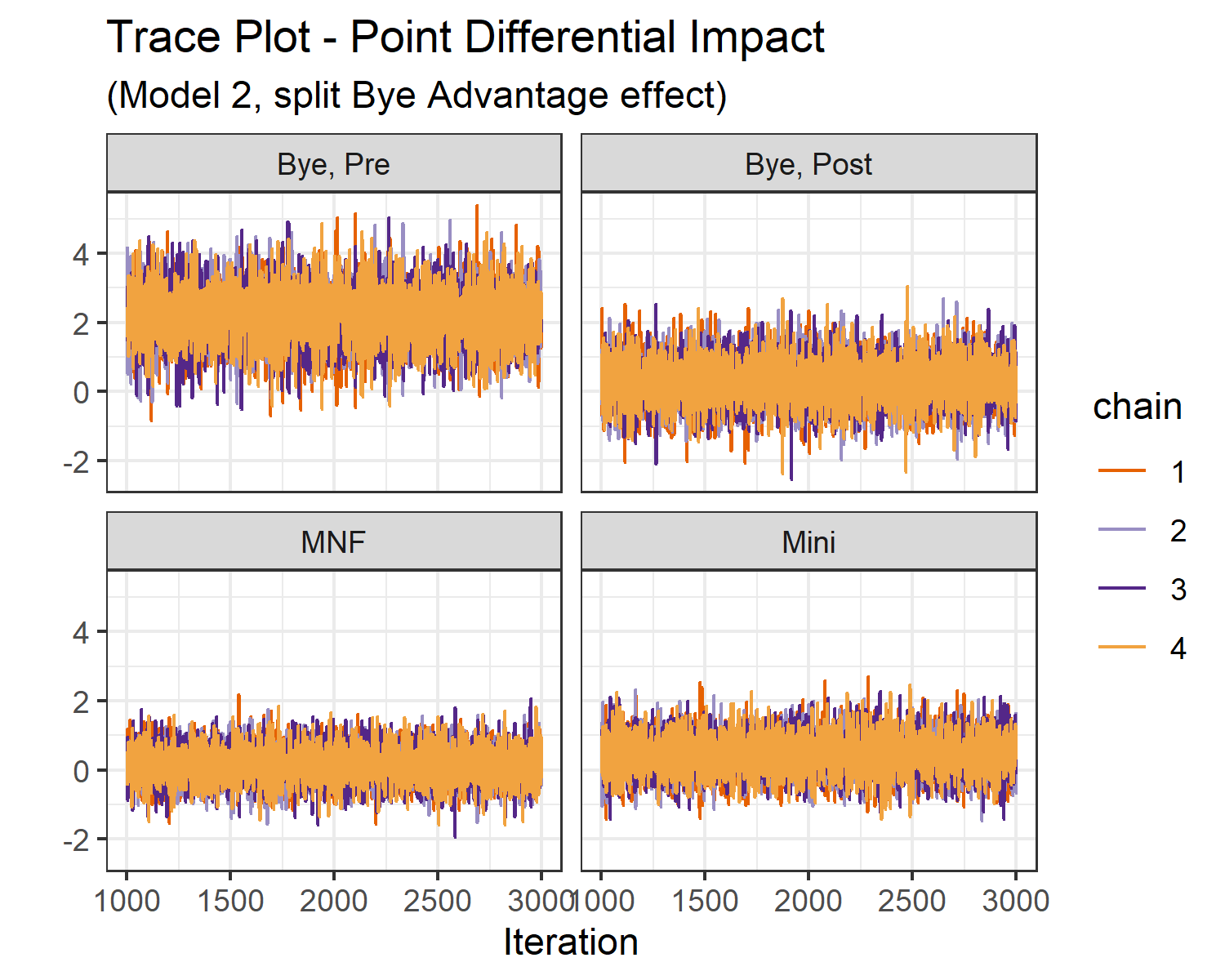}
    \caption{Trace plot for each rest parameter of interest for Model \ref{eq:model2}. There is no evidence of a lack of convergence.}.
    \label{fig:model_2_trace}
\end{figure}

\begin{figure}[hbt!] 
    \centering
    \includegraphics[width=\textwidth]{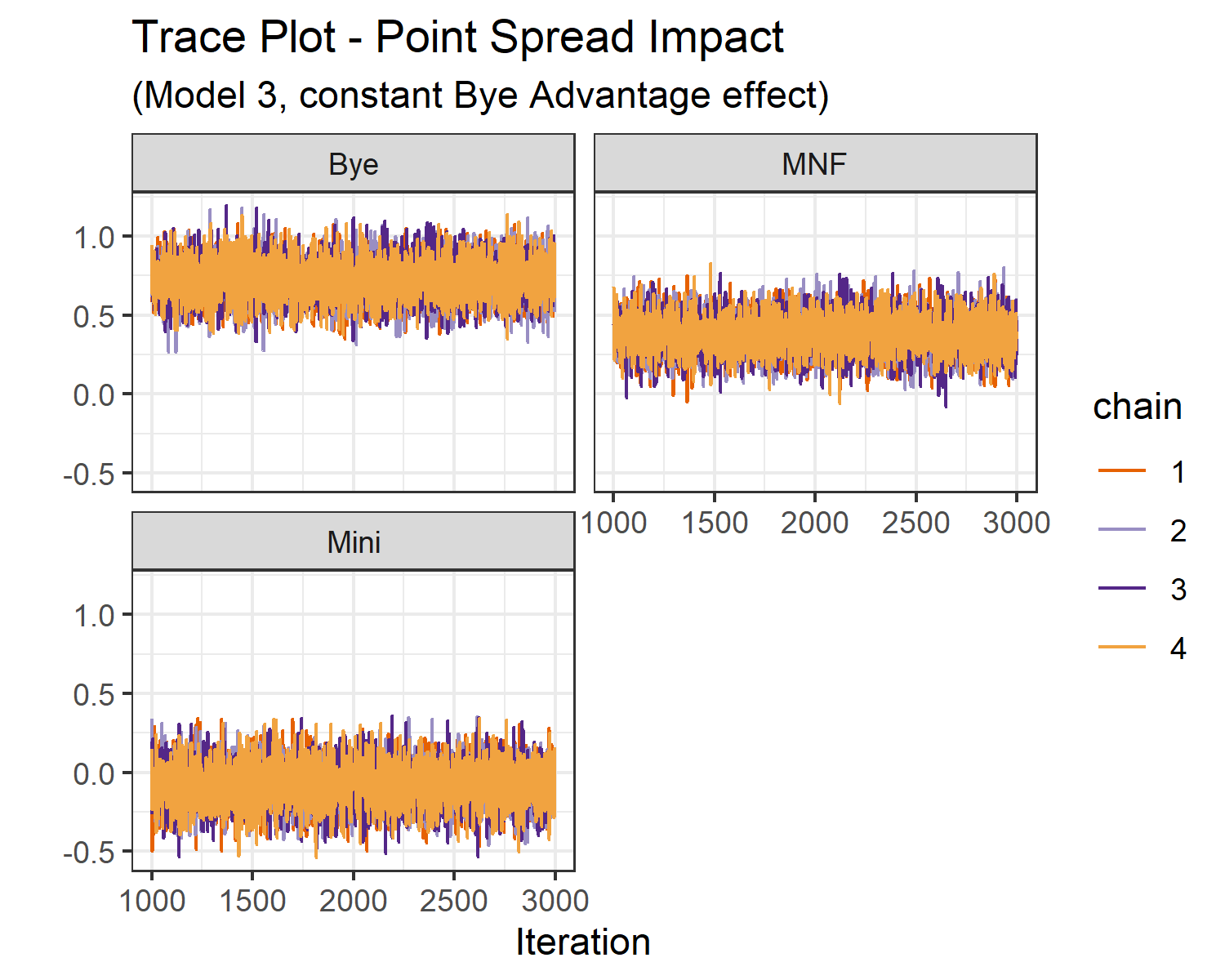}
    \caption{Trace plot for each rest parameter of interest for Model \ref{eq:model3}. There is no evidence of a lack of convergence.}.
    \label{fig:model_3_trace}
\end{figure}

\begin{figure}[hbt!] 
    \centering
    \includegraphics[width=\textwidth]{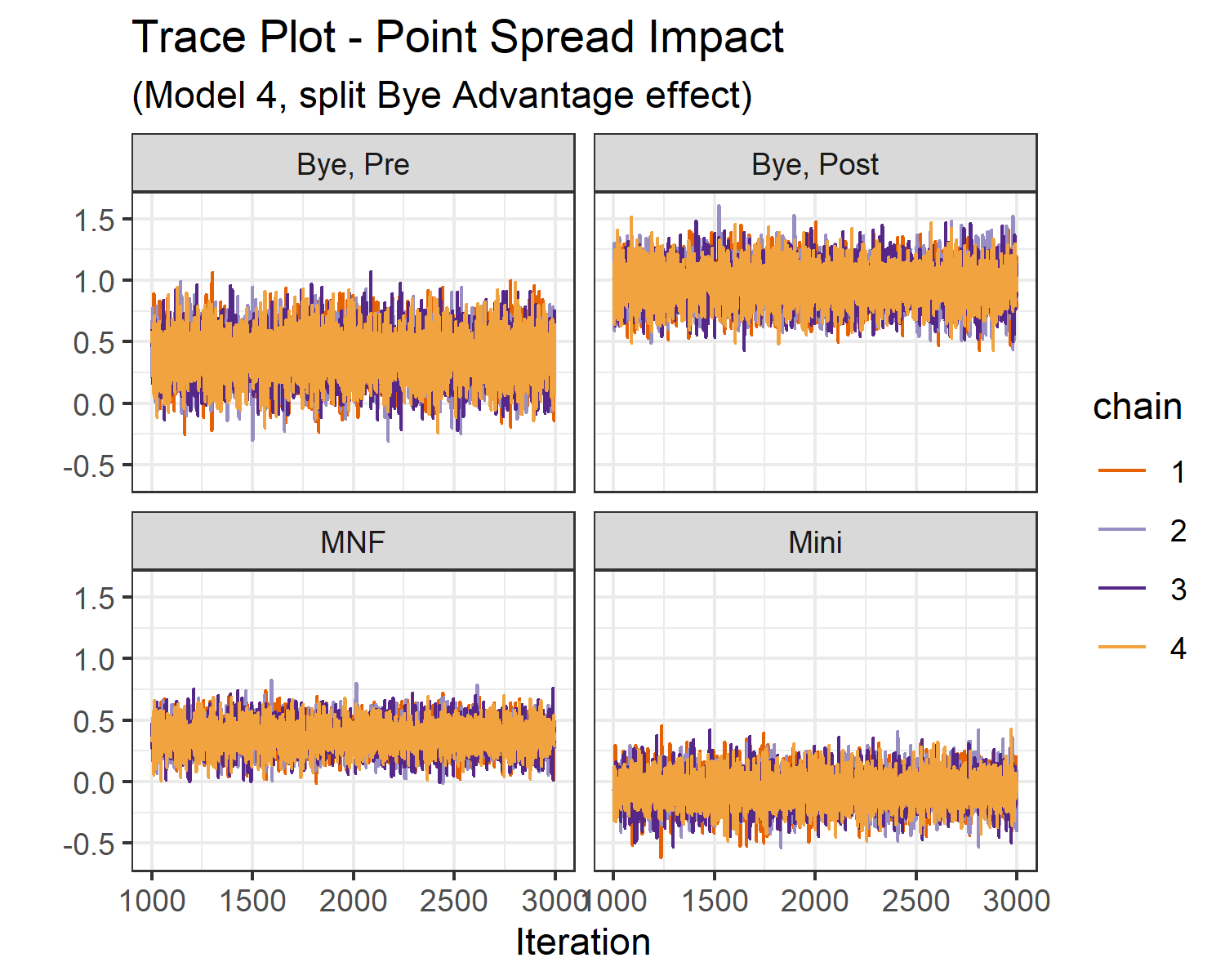}
    \caption{Trace plot for each rest parameter of interest for Model \ref{eq:model4}. There is no evidence of a lack of convergence.}.
    \label{fig:model_4_trace}
\end{figure}

\begin{table}[hbt!] 
\centering
\begin{tabular}{rrlll}
  \hline
Season & Week & Home Team & Away Team & Drop Reason \\ 
  \hline
2010 & 16 & PHI & MIN & Rescheduled \\ 
  2010 & 17 & DET & MIN & Game After Rescheduled \\ 
  2010 & 17 & PHI & DAL & Game After Rescheduled \\ 
  2020 & 4 & KC & NE & Rescheduled \\ 
  2020 & 5 & KC & LV & Game After Rescheduled \\ 
  2020 & 5 & TEN & BUF & Rescheduled \\ 
  2020 & 6 & MIA & NYJ & Rescheduled \\ 
  2020 & 6 & TEN & HOU & Game After Rescheduled \\ 
  2020 & 6 & BUF & KC & Rescheduled \\ 
  2020 & 6 & NE & DEN & Rescheduled \\ 
  2020 & 7 & LAC & JAX & Rescheduled \\ 
  2020 & 7 & DEN & KC & Game After Rescheduled \\ 
  2020 & 7 & NE & SF & Game After Rescheduled \\ 
  2020 & 7 & NYJ & BUF & Game After Rescheduled \\ 
  2020 & 7 & TEN & PIT & Rescheduled \\ 
  2020 & 8 & DEN & LAC & Rescheduled \\ 
  2020 & 8 & CIN & TEN & Game After Rescheduled \\ 
  2020 & 8 & MIA & LA & Game After Rescheduled \\ 
  2020 & 8 & BAL & PIT & Rescheduled \\ 
  2020 & 9 & DAL & PIT & Game After Rescheduled \\ 
  2020 & 9 & LAC & LV & Game After Rescheduled \\ 
  2020 & 9 & ATL & DEN & Game After Rescheduled \\ 
  2020 & 9 & IND & BAL & Game After Rescheduled \\ 
  2020 & 9 & JAX & HOU & Game After Rescheduled \\ 
  2020 & 10 & MIA & LAC & Rescheduled \\ 
  2020 & 11 & DEN & MIA & Rescheduled \\ 
  2020 & 11 & LAC & NYJ & Rescheduled \\ 
  2020 & 12 & DEN & NO & Game After Rescheduled \\ 
  2020 & 12 & BUF & LAC & Game After Rescheduled \\ 
  2020 & 12 & NYJ & MIA & Game After Rescheduled \\ 
  2020 & 12 & PIT & BAL & Rescheduled \\ 
  2020 & 13 & PIT & WAS & Rescheduled \\ 
  2020 & 13 & BAL & DAL & Rescheduled \\ 
  2020 & 14 & CLE & BAL & Game After Rescheduled \\ 
  2020 & 14 & BUF & PIT & Game After Rescheduled \\ 
  2020 & 14 & SF & WAS & Game After Rescheduled \\ 
  2020 & 14 & CIN & DAL & Game After Rescheduled \\ 
  2021 & 15 & TB & NO & Game After Rescheduled \\ 
  2021 & 15 & LA & SEA & Rescheduled \\ 
  2021 & 15 & JAX & HOU & Game After Rescheduled \\ 
  2021 & 15 & PIT & TEN & Game After Rescheduled \\ 
  2021 & 15 & PHI & WAS & Rescheduled \\ 
  2021 & 15 & CLE & LV & Rescheduled \\ 
  2021 & 16 & PHI & NYG & Game After Rescheduled \\ 
  2021 & 16 & LV & DEN & Game After Rescheduled \\ 
  2021 & 16 & MIN & LA & Game After Rescheduled \\ 
  2021 & 16 & DAL & WAS & Game After Rescheduled \\ 
  2021 & 16 & SEA & CHI & Game After Rescheduled \\ 
  2021 & 16 & GB & CLE & Game After Rescheduled \\ 
  2022 & 17 & CIN & BUF & Cancelled \\ 
  2022 & 18 & BUF & NE & Game After Cancelled \\ 
  2022 & 18 & CIN & BAL & Game After Cancelled \\ 
   \hline
\end{tabular}
\caption{Games dropped from our analysis, as well as the explanation. Most games dropped were due to schedule changes during the Covid-19 pandemic. These games, as well as each of the ensuing games for each team, do not reflect typical NFL practice and rest schedules.}
\label{tab:gamesdropped}
\end{table}

\end{document}